\documentclass[]{elsarticle}
\usepackage{graphicx}
\usepackage{amssymb}\usepackage{url}
\usepackage{lscape}
\usepackage{natbib}
\usepackage{url}
\bibpunct{(}{)}{;}{a}{}{,}
\include{bib_alias}

\journal{Icarus}

\begin{document}

\begin{frontmatter}


\title{CoRoT: harvest of the exoplanet program}

\author[lam]{Claire Moutou \corref{cor1}}
\ead{Claire.Moutou@oamp.fr}
\cortext[cor1]{Corresponding author.}
\author[lam]{Magali Deleuil}
\author[oca]{Tristan Guillot}
\author[lesia]{Annie Baglin}
\author[ias]{Pascal Bord\'e}
\author[iap]{Francois Bouchy}
\author[dlr]{Juan Cabrera}
\author[dlr]{Szil\'ard Csizmadia}
\author[iac]{Hans J. Deeg}
\author{and the CoRoT Exoplanet Science Team}
\address[lam]{Aix Marseille University, CNRS, LAM (Laboratoire d'Astrophysique de Marseille) UMR 7326, 13388 Marseille cedex 13, France }
\address[oca]{Observatoire de la C\^ote d'Azur, Laboratoire Lagrange, BP 4229, 06304 Nice cedex 4, France }
\address[lesia]{LESIA, Observatoire de Paris, Place Jules Janssen, 92195 Meudon cedex, France }
\address[ias]{Institut d'Astrophysique Spatiale, Universit\'e Paris 11, CNRS, UMR 8617, b\^at. 121, 91405 Orsay, France}
\address[iap]{Institut d'Astrophysique de Paris , 98bis boulevard Arago, 75014 Paris, France}
\address[dlr]{Institute of Planetary Research, German Aerospace Center, Rutherfordstrasse 2, 12489 Berlin, Germany }
\address[iac]{Instituto de Astrofisica de Canarias, E-38205 La Laguna, and Universidad de La Laguna, Dept. de Astrof\'\i sica, E-38200 La Laguna,Tenerife, Spain}

\date{Draft \today}


\begin{abstract}

{{One of the objectives of the CoRoT mission is the search for transiting extrasolar planets using high-precision photometry, and the accurate characterization of their fundamental parameters.}
{The CoRoT satellite consecutively observes crowded stellar fields since February 2007, in high-cadence precise photometry; periodic eclipses are detected and analysed in the stellar light curves. Then complementary observations using ground-based facilities allows establishing the nature of the transiting body and its mass.}
 {CoRoT has acquired more than 163\,000 light curves and detected about 500 planet candidates. A fraction of them (5\%) are confirmed planets whose masses are independently measured. Main highlights of the CoRoT discoveries are: i) the variety of internal structures in close-in giant planets, ii) the characterisation of the first known transiting rocky planet, CoRoT-7 b, iii) multiple constraints on the formation, evolution, role of tides in planetary systems. }
  {}}
\end{abstract}


\end{frontmatter}

\section{Introduction}
Extrasolar planets whose orbit is aligned with their central star's line of sight are transiting planets. Their observation, when photometry and spectroscopy are combined, allows the precise determination of  fundamental parameters such as the planet radius, the planet true mass, and all orbital elements. Comparative planetology has risen fast with the large number of results obtained from ground-based photometric surveys, especially the multi-site, high-cadence SuperWASP and HAT networks  \citep{cameron,bakos}. Photometry from space is less limited by atmospheric variations, Earth rotation and duty cycle, and gives access to transiting exoplanets of longer period and/or shallower transit depths. One of the science objectives of the pioneering space mission CoRoT is the search for transiting planets from space.\\

The CoRoT (Convection, ROtation and Transits) satellite is a CNES mission with contributions from Austria, Belgium, Brazil, ESA, Germany and Spain \citep{baglin03}. It was launched six years ago and has been observing since February 2007. It was the first mission in space aiming at exoplanet discoveries, two and a half years before the launch of the more ambitious Kepler mission \citep{borucki}. The second objective of CoRoT is asteroseismology, carried on a separate channel for stars brighter than V=8. The Exoplanet program of CoRoT consists of i) the analysis of about 6000 light curves per detector and pointing, ii) the detection of transiting candidates and iii) the follow-up observations to establish the  nature of the candidates and finalise the characterisation of the confirmed planets. The space photometric missions CoRoT and Kepler have been changing our view and understanding of the exoplanet population, by providing unprecedented precision and variety in the fundamental parameters of the planetary systems. More than 180 new transiting planets have been announced since 2007, and about half of them from these two space missions, including multiple planetary systems and smaller planets with accurately determined parameters. In this review of the CoRoT Exoplanet program, we will restate the mission characteristics (section 2), describe the properties of the candidates (section 3) and focus on some specific scientific results (section 4). Finally, we discuss the CoRoT results and statistics in the context of other exoplanet surveys (section 5) and conclude on the prospective work on transiting planets.

\section{Mission characteristics}
CoRoT was launched in December 2006, on a polar orbit at 900 km altitude. The primary mirror of the telescope is 27 cm, giving access to precise relative photometry of stars in the magnitude range $r$ = 11-16.5. CoRoT started operations in February 2007. Observations were interrupted, since then, only a few days per year to change the pointing. From its low polar orbit, CoRoT can alternately observe a region centered in [6h50, 0$^\circ$] equatorial coordinates -hereafter called the "Anticenter", near Monoceros constellation- from October to March and a region centered in [18h50, 0$^\circ$] coordinates -the "Center", near Aquila constellation- from March to October. In the exoplanet field of view, the observations are carried out with the following characteristics: up to 5640 stars per detector (1.40$^\circ$x1.40$^\circ$) are observed. During the two first years of operations, CoRoT had two detectors in the exoplanet channel; one of them failed in March 2009. 

\begin{figure}
  \begin{center}
     \includegraphics[scale=0.45]{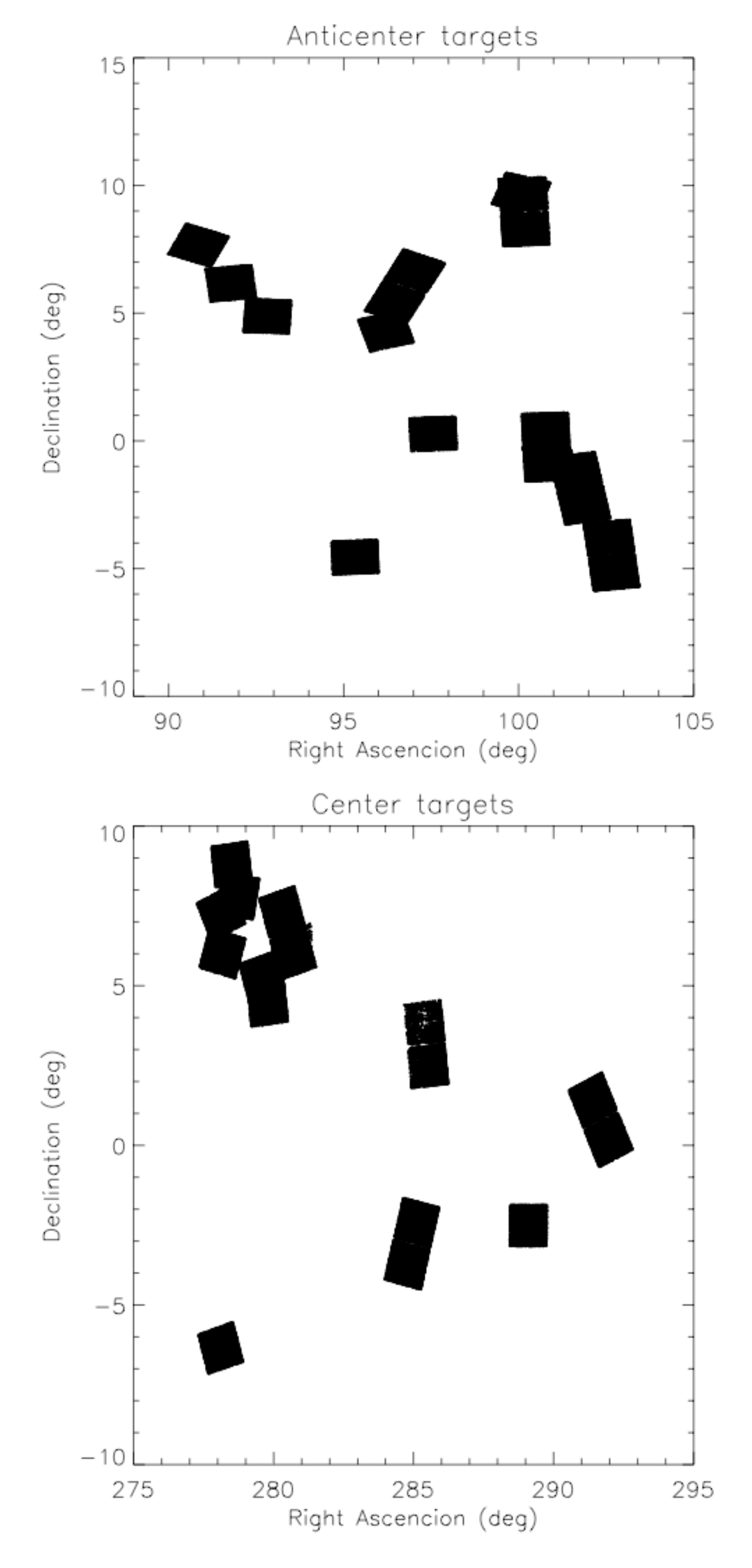}
    \caption{The location of CoRoT observed targets towards the Galactic Anticenter (top) and towards the Galactic Center (bottom).}
    \label{fig:fields}
  \end{center}
\end{figure}

Figure \ref{fig:fields} shows the location of CoRoT targets observed in the exoplanet field from February 2007 to October 2012 in equatorial coordinates. CoRoT typically observes 3 to 4 different fields per year, and thus may potentially increase the target sample by $\sim$20\,000 additional stars each year. In practice, there has been some overlap between runs, and several thousands stars have been observed in two or three runs. In total, 163\,664 stars  have been observed from February 2007 to October 2012, and the total area observed by CoRoT covers about 58 square degrees during 26 different runs. The duration of the runs depends on the leading science objective, taking into account the requirements in stellar physics; the average run duration is 78 days, the shortest run had a duration of 21 days  and the longest one, 152 days. The duty cycle slightly exceeds 91\%. \\

For one third of the observed targets among the brightest dwarf stars, three light curves are computed onboard, which contain different color and spatial information. The three-colour light curves can be analysed in combination with the catalogue of stellar contaminants, to help identifying blend scenarios and calculate the precise transit contamination factor. Forty stars are observed in 40$\times$40 pixel imagettes that are fully transferred to Earth every 32 s; these ones receive a specific data treatment and signal extraction is optimised \citep{barros}. The other targets are observed in a monochromatic mode, and the spatial information is limited to the $\sim$ 20 arcsec aperture where the photometry is computed. All target properties, including multi-color catalogue photometry, stellar classification, variability properties, characteristics of neighbour stars, and observation characteristics, are available in the CoRoT data base http://cesam.oamp.fr/exodat \citep{deleuil09}. \\
 
The noise properties and detection threshold in the light curves of the exoplanet channel are described in several papers \citep{aigrain09,mazeh09,cabrera09,borde,bonomo12}. The early noise analyses show a performance of about 7.10$^{-4}$ in 2h timescale for a $R = 15$ magnitude star. Most light curves acquired  during the first year of operations have a noise level exceeding the photon noise by a factor 2, with a low level of correlated noise. In \citet{bonomo12}, the use of near-threshold simulated transits on CoRoT light curves allowed an accurate description of the instrumental limits. CoRoT light curves are affected by jitter noise and sudden jumps due to hot pixels, and these characteristics require the optimisation of the filtering methods to lower their effects in the transit detection process. In summary, about 60\% of simulated transits due to planets of 2-4 Earth radii are detected around G and K stars brighter than 14, and this number decreases to 10\% when planets of 1.3 to 2 Earth radii are concerned.
 
The evolution of noise in the instrument behaves as slowly as expected: after 1500 days in operations, the number of pixels exceeding the value of 1000 electrons (so-called hot pixels) amounts to 1.5\% of the total number of pixels, and the sensitivity decrease follows a linear slope of 6.4 $\cdot$ 10$^{-5}$ electrons per day (M. Auvergne, priv. comm.). In the most recent light curves, the noise may reach three times the photon noise; if most light curves evolve smoothly, there is some residual scatter in the global behaviour. This scatter may be due to the background correction that needs to be further optimised  to take into account an evolving gradient on the detectors, and is currently under investigation.
 
The transit detection and the ranking of the planetary candidates result from a coordinated analysis, involving several detrending methods \citep[eg,][]{aigrain04,bonomo12,borde07,cabrera12,carpano08,grziwa,mazeh09,ofir,tingley11b}. Most candidates are detected by all algorithms, and only a small fraction of additional candidates -- usually shallow transits, light curves with a low number of events, and/or photometrically active stars -- are found by only one algorithm.  This assessment was already known from pre-launch simulations \citep{moutou05} and only emphasizes the impact of noise, stellar variability and filter methods on transit detection.

\section{Candidates}
A comprehensive work on all transiting events discovered by CoRoT is published in \citet{mega}, after analyses of the first transiting candidates were reported in earlier works \citep{carpano09,moutou09,cabrera09,erikson,cavarroc,carone}. Transit events have been detected in more than 3900 light curves, 86\% of them from eclipsing binaries. Up to September 2012, 530 candidates were flagged as planetary like, with depths ranging from 0.02 to 5\%  and periods ranging from 0.5 to 95 days. The distribution of periods and depths shown in \citet{mega}  has peak values of 5 days for the orbital period and 0.2\% for the transit depth (see also Figure \ref{fig:fig3}).\\

\begin{figure}
  \begin{center}
     \includegraphics[scale=0.3]{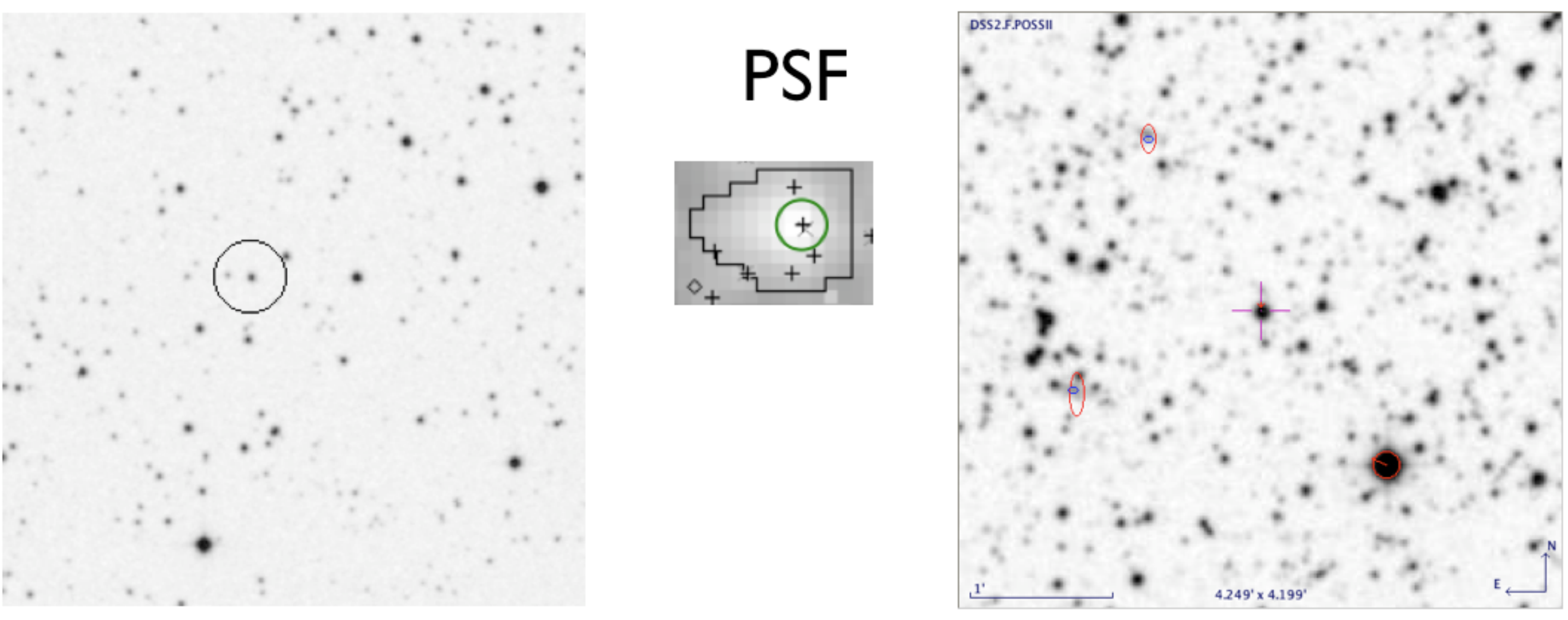}
    \caption{Ground-based image of a typical anticenter (left) or center (right) 3-arcmin field within the CoRoT eyes. The middle image shows the image of a star as seen by CoRoT and the optimal photometric aperture mask (30-arcsec wide) that is assigned to such a PSF. }
    \label{fig:fig2}
  \end{center}
\end{figure}

\begin{figure}
  \begin{center}
     \includegraphics[scale=0.4]{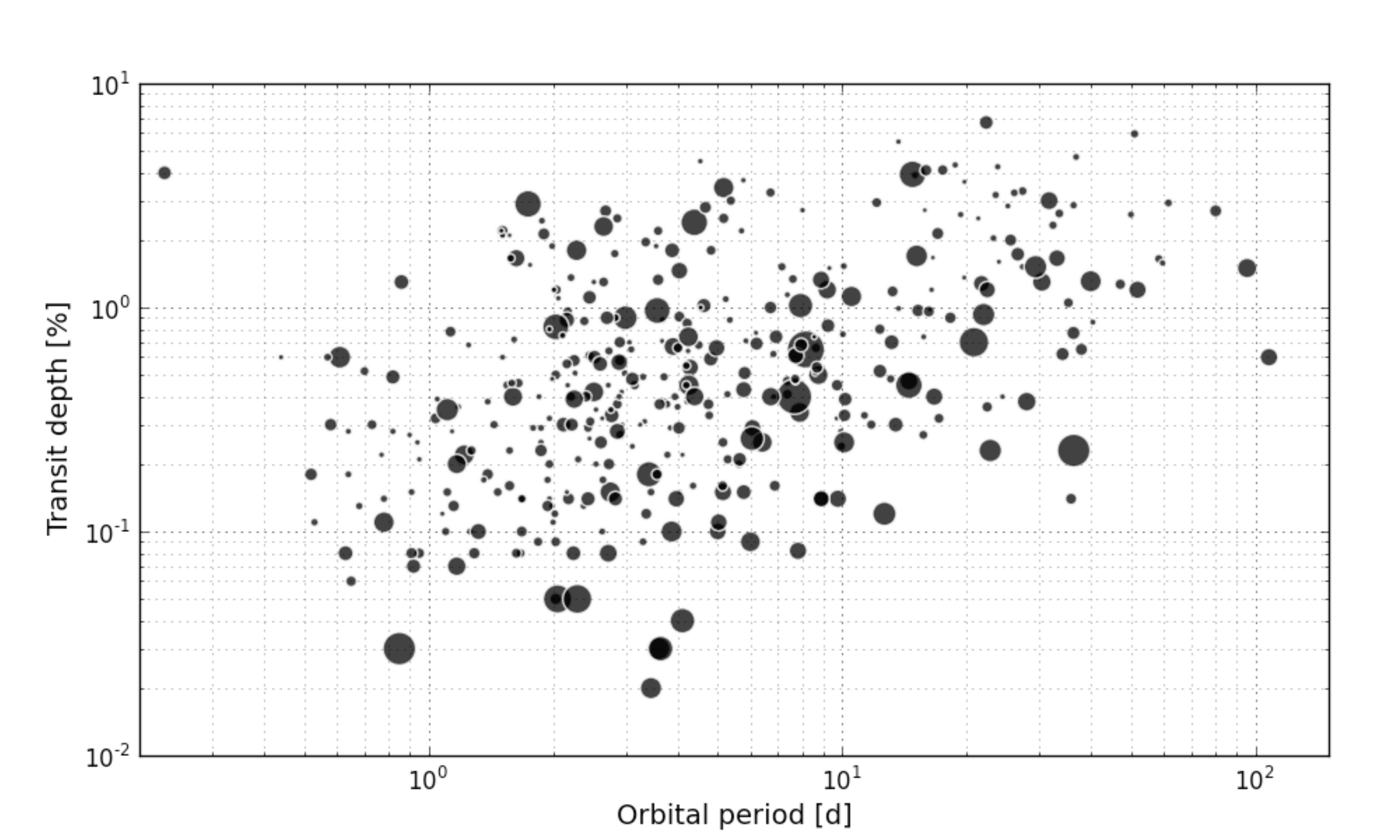}
    \caption{The period and depth of CoRoT planetary candidates (courtesy A. Santerne). The size of the symbols indicates the apparent luminosity of the parent star (small meaning faint).}
    \label{fig:fig3}
  \end{center}
\end{figure}

An intensive effort of complementary ground-based observations has been started as soon as 2007, when the first candidates were detected, with more than 60\% of the detected candidates having some observations performed by one or several among a dozen of different facilities. The first need for complementary observations is illustrated in Figure \ref{fig:fig2}: the stellar field observed by CoRoT is crowded at different levels, depending on the pointing, and the photometric aperture spreads over 5 to 20 arcseconds. The number of contaminating stars per aperture may thus be large, and the light curve properties are not always sufficient to disentangle field binaries from bona fine planetary candidates. 
Follow-up observations mainly consist of i) medium-resolution imaging to exclude neighbour stars as potential eclipsing binaries \citep{deeg09} and ii) high-precision radial-velocity observations to further reject undiluted binaries, to identify some configurations of diluted systems \citep{bouchy09}, and to provide spectral classification of host stars. It is more than 200 nights of 1-2 m class telescopes that have been used for ground-based photometry from European and US observatories, and 350 nights of radial-velocity instruments, mainly on ESO/HARPS and OHP/SOPHIE, with contributions from Keck/HIRES and NOT/FIES facilities. In a few cases, adaptive optics imaging was also performed, in order to search for closer and/or fainter neighbours to target stars \citep{guenther12}, or high-resolution near-infrared spectroscopy \citep{guenther10}. More than 280 candidates were fully resolved by these complementary observations. About 250 candidates remain unsolved, for different or combined reasons: \begin{itemize}
\item the target is fainter than $V=16$ and the transit candidate is not worth an intensive observational effort (low ranking priority) (roughly 20\% of the unsolved cases)
\item the candidate has been observed in radial velocity, but the star rotates fast and it is not possible to set constraint on the mass  (20\%) 
\item the candidate has not been observed early enough, and transit ephemeris have been lost for ground-based imaging confirmation (10\%)
\item the candidate has been extensively observed, but the data are not fully conclusive (ie, without significant signal detected in the radial velocity data set) (50\%).
\end{itemize}

Table \ref{fpp} gives the distribution of the 530 candidates into the 4 categories: i) planets, ii) diluted binaries (solved by ground-based photometry or blend analyses), iii) undiluted binaries (solved by radial-velocity binary detection or refined light curve analyses), and iv) unresolved cases. The proportions are similar in the anticenter and center directions, except a small but significant increase of diluted binaries in the more contaminated center fields. Earlier statistical analyses of false alarms had been reported in \citet{almenara09}.

An interesting class of candidates is the sample of mono-transits, when only one transit event is detected. There are 27 of them identified and well ranked in the CoRoT  candidate list. Their ranking indicates the compatibility between the transit shape and the detection of a unique event (hence, a minimum value for the orbital period and a plausible planetary radius). Few mono-transits have been observed with radial-velocity instruments, and in a few cases only, undiluted binaries were detected. Most of these cases are thus either diluted binaries or long-period planets. The resources in ground-based telescope necessary to resolve such cases, without a constraint on the orbital period, are beyond the capacity of the CoRoT team.

\begin{table}
  \begin{center}
\caption{Number of candidates and distribution of their nature.}
\label{fpp}
\begin{tabular}{lcc}
\hline
Type & Anticenter & Center \\
\hline
number of candidates & 223  & 307  \\
planets (\%) & 6.3& 5.5 \\
diluted binaries (\%)& 17.5 & 20.6\\
undiluted binaries (\%) & 26.0 & 28.0\\
unresolved (\%) & 50.2 & 45.9 \\
\hline
\end{tabular}
  \end{center}
\end{table}

\section{Detection yield}

After 5.5 years of operations, the CoRoT false positive rate (number of non-planet configurations divided by the total number of transiting candidates) is 73$\pm7$\%, for both pointing directions. This percentage is obtained when one assumes that the number of unsolved cases is distributed on planets and diluted binaries as for the solved cases (see Table \ref{fpp}). With a more conservative approach, one can only say that 6\% of the detected candidates turn out to be confirmed planets, with a mass measurement through the radial-velocity detection of the planet signature. The number of confirmed planets is mainly limited by performances of complementary observations, in relation with the faint luminosity of the target stars.\\

Estimating the planet occurrence of the CoRoT mission requires as inputs the number of relevant observed stars, the final number of confirmed planets, the detection yield of the instrument, and corrections for the transit probability. The calculation has been done for hot Jupiters where the detection yield of CoRoT is optimal (assumed 100\%). The result is that 1$\pm$0.3\% hot-Jupiter  planets are detected in the CoRoT-Exoplanet channel stellar sample; the error bar includes the misclassification of stars on which giant planet detection is guaranteed. This is an update with respect to the earlier computation by \citet{guenther12b} from a partial planet list in few Anticenter fields. The occurrence of hot Jupiter planets is compatible between the Center and Anticenter directions. This similarity follows expectations: although the Center and Anticenter directions include  different populations of the Milky Way when a large volume is considered, the dwarf components have the same properties in terms of distance and metallicity distribution \citep{gazzano10}. The frequency of hot Jupiters, known to be affected by the stellar metallicity \citep{gonzalez,santos,johnson}, is therefore expected to be similar in both pointing directions. In addition, the distribution of metallicities of CoRoT and Kepler confirmed planets is also similar (0.012 $\pm$  0.15 for CoRoT planets and 0.012 $\pm$ 0.21 for Kepler planets, from data in www.exoplanets.org), while the occurrence of hot jupiter planets in the Kepler survey is 0.9$\pm$0.1\% \citep{santerne}. In conclusion, the frequencies of hot-jupiter planets found by CoRoT and Kepler are compatible with each other, and are in agreement with the discovery rate of radial-velocity, estimated as 0.89$\pm$0.36 \% \citep{mayor} or in the range 0.8 to 2.8 \% as found by \citet{howard}.\\

Concerning the hot-Neptune planets, one must account for a lower detection yield of CoRoT, which is estimated to be 59 (31)\% for 2-4 Earth radius planets with orbital periods less than 20 days, around stars of maximum magnitude 14.0 (resp. 15.5) \citep{bonomo12}. In such domain, the CoRoT planet yield is too low by a factor two at least, compared to the frequency of Kepler candidates. Several reasons for this discrepancy are proposed in \citet{bonomo12}, such as hypotheses on the estimated stellar radii of Kepler targets, on the Kepler false positive rate, on the stellar classification of CoRoT stars, or a difference in stellar population between both fields; a better understanding of this result is expected, as the resolution of CoRoT and Kepler candidates will make progress.\\

Finally, the huge breakthrough brought by the discovery of several hundreds of multiply-transiting systems by Kepler \citep{latham,lissauer} has no equivalent in the CoRoT mission. CoRoT has found two multiple systems: CoRoT-7 where only the inner planet transits its stars and other low-mass planets are discovered by their radial-velocity signature only  \citep{queloz,hatzes}, and CoRoT-24 where two planets transit the central star \citep{alonso13}.  A couple of other multiply-transiting systems have been discarded by complementary observations. This main difference between both surveys can be understood because of the mission profiles and the properties of observed multiply-transiting systems. Most of these contain small planets, with transit depths in the range 10$^2$ to 10$^3$ ppm, and  a distribution of periods peaking around 5-50 days. This type of transiting events represents challenging detections for CoRoT, with a sparse number of individual transits and a low signal-to-noise ratio per planet candidate. The unique system CoRoT-24 is made of two planets of 2.6 and 4.2 Earth radii and periods of 5.1 and 11.8 days \citep{alonso13}, which properties are similar to the bulge where most Kepler multiply-transiting planets lie. The CoRoT detection yield in this domain is very low, probably less than $\sim$20\%.\\

While most of the Kepler candidates are out of reach for high-precision radial-velocity confirmation of their planetary nature, it is the case for a smaller fraction of CoRoT candidates (about 25\%) because of the difference in detection limits. An alternative route of statistical validation is then required  \citep{torres,fressin,diaz}, which consists of statistically excluding the other possible scenarios featuring a transit, i.e., diluted or undiluted eclipsing binaries. First applications of binary rejection using radial-velocity data to CoRoT candidates were used in the analysis of CoRoT-7 b and CoRoT-8 b \citep{leger, borde}. The "Blender" validation process developed in the context of Kepler candidates has been recently applied to CoRoT-7 b \citep{fressinc7} and showed that the odds ratio between the planet and false-positive scenarios is of the order of 3500, under conservative hypotheses on the planet frequency. In addition to CoRoT-22 \citep{moutou13} and CoRoT-24 b and c \citep{alonso13}, three low-mass planets in the publication process, there are a dozen candidates suitable for planet validation, meaning that all obvious alternative scenarios have been excluded and a sufficient number of additional measurements are available. These promising CoRoT candidates are being analysed with PASTIS, a planet-validation tool using numerical simulations and a Bayesian approach for model comparison \citep{diaz}. For such candidates, however, this is only an upper limit on the mass that is available, and the internal structure of such planets can only be marginally constrained by theory.

\section{Planets}

In this section, we describe the properties of confirmed CoRoT  planets, including only the cases where a significant radial-velocity signature had been detected. At the time of writing, 21 confirmed planets and one brown dwarf are published. A handful of additional planets are to be published soon. Two other systems, one of which including two transiting planets, have been assigned a CoRoT name (CoRoT-22 and CoRoT-24) but their validation as planets is not yet certified by an independent detection of the radial-velocity planet signal, and their case will be discussed in the next section.

\begin{figure}
  \begin{center}
       \includegraphics[scale=0.3]{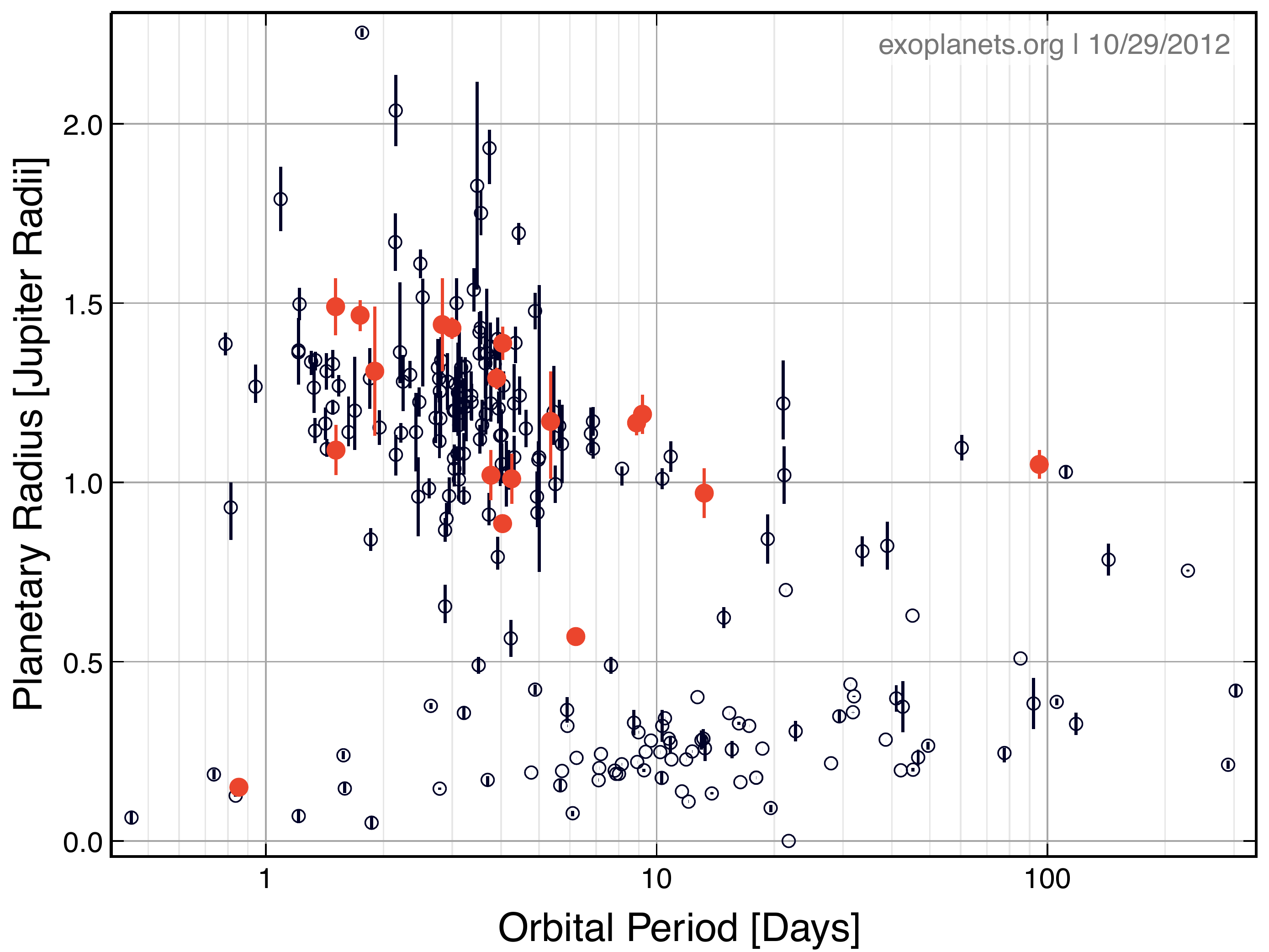}
    \caption{The period and radius distribution of the transiting exoplanets (empty circles) and CoRoT exoplanets (filled red circles).}
    \label{fig:RP}
  \end{center}
\end{figure}

The properties of CoRoT planetary systems are summarized in Table \ref{planets}, where the eccentricities have been updated with recent re-analyses of the radial-velocity data by \citet{husnoo}. The variety of close-in giant planets has been a surprise since the early discovery of the inflated planet HD 209458 b \citep{charbonneau, henry}, especially the variety of radii for a given planet mass. The planet radius covers a decade, from 0.149  (CoRoT-7 b) to 1.49 Jupiter radius (CoRoT-1 b, \citet{barge}). With the additional determination of planet masses, the mean density of CoRoT planets ranges from 0.2 g.cm$^{-3}$ (CoRoT-5 b, \citet{rauer}) to 26  g.cm$^{-3}$ (CoRoT-3 b, \citet{deleuil3}). The periods of CoRoT confirmed planets range from 0.85 day (CoRoT-7 b, \citet{leger}) to 95.27 days (CoRoT-9 b, \citet{deeg}). Figure \ref{fig:RP} shows the distribution of radii and periods of CoRoT planets and other transiting planets. In between rocky and giant planets lies the well-characterized CoRoT-8\,b \citep{borde}, which has a density similar to Neptune, an estimated mass of heavy elements comparable to HD\,149026 but a much smaller hydrogren-helium envelope.
Examples of  topics where CoRoT results contributed to answer some scientific questions are given in the next sections.\\

\subsection{Telluric exoplanets}

Although the scientific community has always been confident about the existence (even abundance) of telluric planets outside the solar system, the detection of the first one having a mean density similar to the Earth density has been a major advance in the field of comparative planetology. CoRoT-7 b has been the first detected rocky planet (see properties on Table \ref{planets}), and has been quickly followed by similar discoveries: the twin planet Kepler-10\,b by \citet{batalha}, and the fluffy super-Earthes GJ 1214\,b by \citet{charbonneau09} and 55 Cnc\,e by \citet{winn}. The class of hot telluric super-Earth planets has triggered new theoretical developments, focusing on the impact on the planet of the extreme environmental conditions, due to their proximity to the host star (only 4.5 stellar radii for CoRoT-7 b). Planet erosion could be large due to UV and particle flux, allowing the loss of 4 to 10 Earth mass at the age of the system \citep{poppenhaeger}. The planet rotation is likely synchronised with its orbital motion, which implies a huge temperature gradient from the night side (50-75K) to the day side (more than 2500K); for this reason, the surface on the day-side is expected to be made of molted rocks, and \citet{leger11} have proposed to name such planets "Lava-ocean planets". In this model, the atmosphere is made of rocky vapours at low pressure  \citep{castan}. Models of the internal structure of CoRoT-7 b have also been developed, in the difficult context of unstable planetary parameters: due to the star activity, the mass of the planet and also the radius of the star have been revised several times \citep{hatzes,ferraz,boisse,pont,bruntt}, with a strong impact on the predicted planet composition \citep{valencia}. The last result points to a planet enriched in iron with a bulk composition similar to Mercury's; it would have mantle convection and thus would generate magnetic field and additional heating to the planet's interior \citep{wagner}. Finally, the properties of CoRoT-7 b most probably makes it an unstable planet as far as libration-driven elliptical instability in synchronized bodies is concerned; this instability may have a strong impact on the heat fluctuation and magnetic induction \citep{cebron}.

\subsection{The inflation problem \& the compositions of giant planets}

A fraction of the giant planets of the CoRoT sample are significantly larger than would be expected from the evolution of a hydrogen-helium planet. This is a classical problem \citep{bodenheimer,guillot02} which is quantified by calculating the {\it radius anomaly}, defined as the difference between the measured radius and that calculated for a solar-composition planet with no core of the same mass, age and equilibrium temperature \citep{guillot06}. This radius anomaly expressed as a fraction of the measured planetary radius, $\delta_ {\rm anomaly}$, is provided for the 20 giant planets of the CoRoT sample in Table~\ref{tab:interior}. Seven planets (CoRoT-1, 2, 5, 6, 11, 12, 19) have significant positive values of $\delta_ {\rm anomaly}$ between 10 and 20\%, indicating that the inflation mechanism is widespread, in line with what found for the ensemble of giant exoplanets \citep{guillot08,laughlin11}. Generally, the problem can be solved by invoking for all of these planets  that a non-radiative process leads to the conversion of a small fraction of the stellar absorbed irradiation (of order 1\%) to kinetic energy and its dissipation at greater depths \citep{guillot02,batygin10}. 

\begin{table}
\caption{Properties of the CoRoT giant planets: Equilibrium temperature, radius anomaly and metal content, for dissipation of $0.125\%$ (a) or $0.25\%$ (b) of the incoming stellar irradiation flux.}\label{tab:interior}
\begin{tabular}{lccccc}
\hline
Name & $M_{\rm p}$ $\rm[M_\oplus]$ & $T_{\rm eq}$ [K] & $\delta_{\rm anomaly}$ & $M_Z^{\rm (a)}$  $\rm[M_\oplus]$  & $M_Z^{\rm (b)}$  $\rm[M_\oplus]$ \\ \hline
CoRoT-1b & 327(38) & 1900(66) & $0.171(65)$ & $-28(27)$ & $-12(26)$ \\
CoRoT-2b & 1052(51) & 1528(137) & $0.120(27)$ & $-181(41)$ & $-166(38)$ \\
CoRoT-3b & 6884(25) & 1688(70) & $-0.011(70)$ &$^a$ &  $^a$ \\
CoRoT-4b & 229(25) & 1074(19) & $0.024(54)$ & $12(14)$ & $20(14)$ \\
CoRoT-5b & 148(13) & 1438(39) & $0.149(38)$ & $22(6)$ & $27(6)$ \\
CoRoT-6b & 941(108) & 1017(21) & $0.051(30)$ & $-47(39)$ & $-31(38)$ \\
CoRoT-8b & 70(10) & 854(24) & $-0.940(66)$ & $39(0)$ & $39(0)$ \\
CoRoT-9b & 263(4) & 411(9) & $0.008(25)$ & $-1(9)$ & $1(9)$ \\
CoRoT-10b & 874(51) & 670(31) & $-0.141(74)$ & $148(72)$ & $147(70)$ \\
CoRoT-11b & 741(108) & 1732(51) & $0.163(23)$ & $-86(22)$ & $-56(21)$ \\
CoRoT-12b & 292(22) & 1444(81) & $0.203(95)$ & $-23(36)$ & $-11(35)$ \\
CoRoT-13b & 416(21) & 1275(38) & $-0.344(47)$ & $179(11)$ & $185(9)$ \\
CoRoT-14b & 2416(191) & 1953(96) & $-0.085(76)$ & $373(170)$ & $432(147)$ \\
CoRoT-16b & 170(16) & 1195(71) & $0.042(46)$ & $28(9)$ & $34(9)$ \\
CoRoT-17b & 782(16) & 1553(78) & $-0.089(50)$ & $194(60)$ & $221(58)$ \\
CoRoT-18b & 1103(121) & 1527(127) & $0.043(142)$ & $-46(202)$ & $-32(197)$ \\
CoRoT-19b & 353(19) & 1660(39) & $0.100(24)$ & $27(11)$ & $42(11)$ \\
CoRoT-20b & 1348(73) & 954(38) & $-0.402(57)$ & $508(54)$ & $486(55)$ \\
CoRoT-21b & 718(99) & 2048(143) & $0.074(111)$ & $83(103)$ & $95(85)$ \\
CoRoT-23b & 890(95) & 1642(119) & $-0.073(125)$ & $163(121)$ & $188(114)$ \\
\hline\hline
\end{tabular}
{\\
The parentheses indicate the uncertainty on the last digits.\\
$^a$ Because of the large mass of CoRoT-3b, models with a core were not calculated.}
\end{table}

However, one of the CoRoT planets is found to resist standard explanations so-far: CoRoT-2b has a radius that is not amongst the largest so far, but it is massive, and explaining its size thus requires a considerable amount of additional energy \citep{alonso09a, gillon10a}, of about 25\% of the absorbed solar luminosity if this is a long-term feature \citep{guillot11}. Interestingly, CoRoT-2 shows all the signs of being young (see next section), so that one possibility may be that the planet was impacted by a Saturn-mass planet, or was circularized from an initially large eccentricity just 10 to 30 Myr ago \citep{guillot11}. An alternative explanation could be the dissipation due to the hydrodynamical instability elliptically deformed by tides, as described in \citet{cebronb}, if the eccentricity of the orbit does not equal zero exactly.

Once a prescription for the missing physics is assumed, the mass of heavy-elements in the giant planets may be calculated \citep{guillot06}: the presence of additional heavy elements either as a central core or inside the envelope generally leads to a shrinking of the planet compared to solar-composition models and naturally explains the negative $\delta_{\rm anomaly}$ values. For simplicity, this is done by assuming that a fraction of the incoming energy is dissipated at the planet's center, and that all the heavy elements are embedded as a central core. Table~\ref{tab:interior} provides the resulting $M_Z$ values in two cases, for the dissipation of $0.125\%$ or $0.25\%$ of the incoming stellar irradiation flux $\sigma T_{\rm eq}^4$, respectively. (Unphysical negative values of $M_Z$ are obtained from a linear extrapolation of the results for no core mass and for a 20\,M$_\oplus$ core mass). With the dissipation of only $0.25\%$ of the incoming energy at the center, all CoRoT planets are found to have positive core masses within error bars, with the exception of CoRoT-2\,b and CoRoT-11\,b. These may however require special explanations, as discussed in the next sections. 

\begin{figure}
  \resizebox{\hsize}{!}{\includegraphics{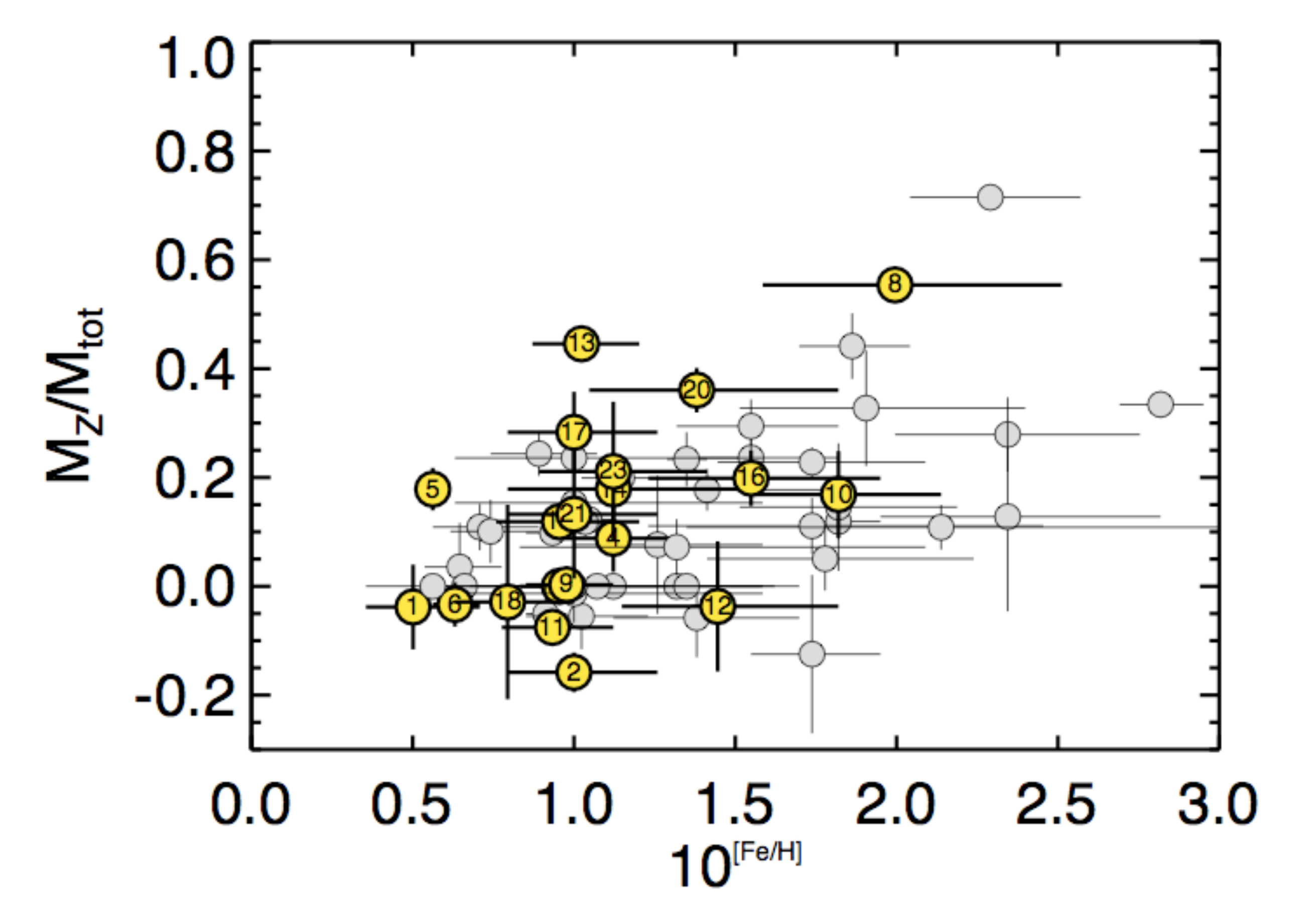}}
  \caption{Mass fraction of heavy elements in the planets as a function of the metal content of their parent star expressed in solar units (e.g. $10^{\rm [Fe/H]}=3$ implies that the star contains three times more iron than our Sun). The evolution model assumes that 0.25\% of the incoming irradiation flux is dissipated at the planet's center. Circles which are labelled 1 to 23 correspond to the CoRoT giant planets (see table~\ref{tab:interior}). Grey symbols correspond to a subset of known transiting systems  \citep{guillot08,laughlin11}. }
  \label{fig:correlation}
\end{figure}

When plotted against stellar metallicity as in Figure~\ref{fig:correlation}, the giant planets in the CoRoT sample appear to have compositions that are correlated with that of their parent stars, confirming the trend observed for many planets \citep{guillot06,burrows07,guillot08,laughlin11}, and for non-irradiated planets \citep{miller11}. The correlation coefficient between $10^{\rm [Fe/H]}$ and $M_Z^{\rm b}/M_{\rm p}$ for the CoRoT planets is $r=0.52$ and a Spearman's test indicate that its statistical significance is $2.3\sigma$. When including other planets as in \citet{guillot08}, the value of $r$ remains the same, but the significance increases to $3.5\sigma$. 

However, some of the CoRoT planets have small radii for large masses and require masses of heavy elements in excess of what is usually envisioned by formation models. This is the case of CoRoT-10\,b, 13\,b, 14\,b, 17\,b, 20\,b and 23\,b which all require core masses in excess of $70\rm\,M_\oplus$ to explain their small size. Both the large amounts required and the relatively high fraction of these superdense planets is surprising. It appears to be larger than in the general sample, which may be attributed to the fact that these planets are generally smaller and more difficult to detect from the ground\footnote {An alternative explanation could be to invoke the blending flux of background stars. However, the mean radius anomaly of the "overdense" planets being of order 19\%, this implies that this explanation would require that for these, the contamination by background stars should have been underestimated by about 34\%. Given the follow-up observations at high spatial resolution led for each target down to magnitude V=19, this explanation is regarded as very unlikely.} On the theoretical standpoint, for large $M_Z$ values, the assumption that all the heavy elements are in the core and the very high central pressures both probably imply that $M_Z$ is significantly overestimated \citep[see][]{deleuil20, baraffe98}.  In any case, further work both on the internal structure of these planets and their formation is required.

\subsection{The young planets}

\begin{figure}
  \resizebox{\hsize}{!}{\includegraphics{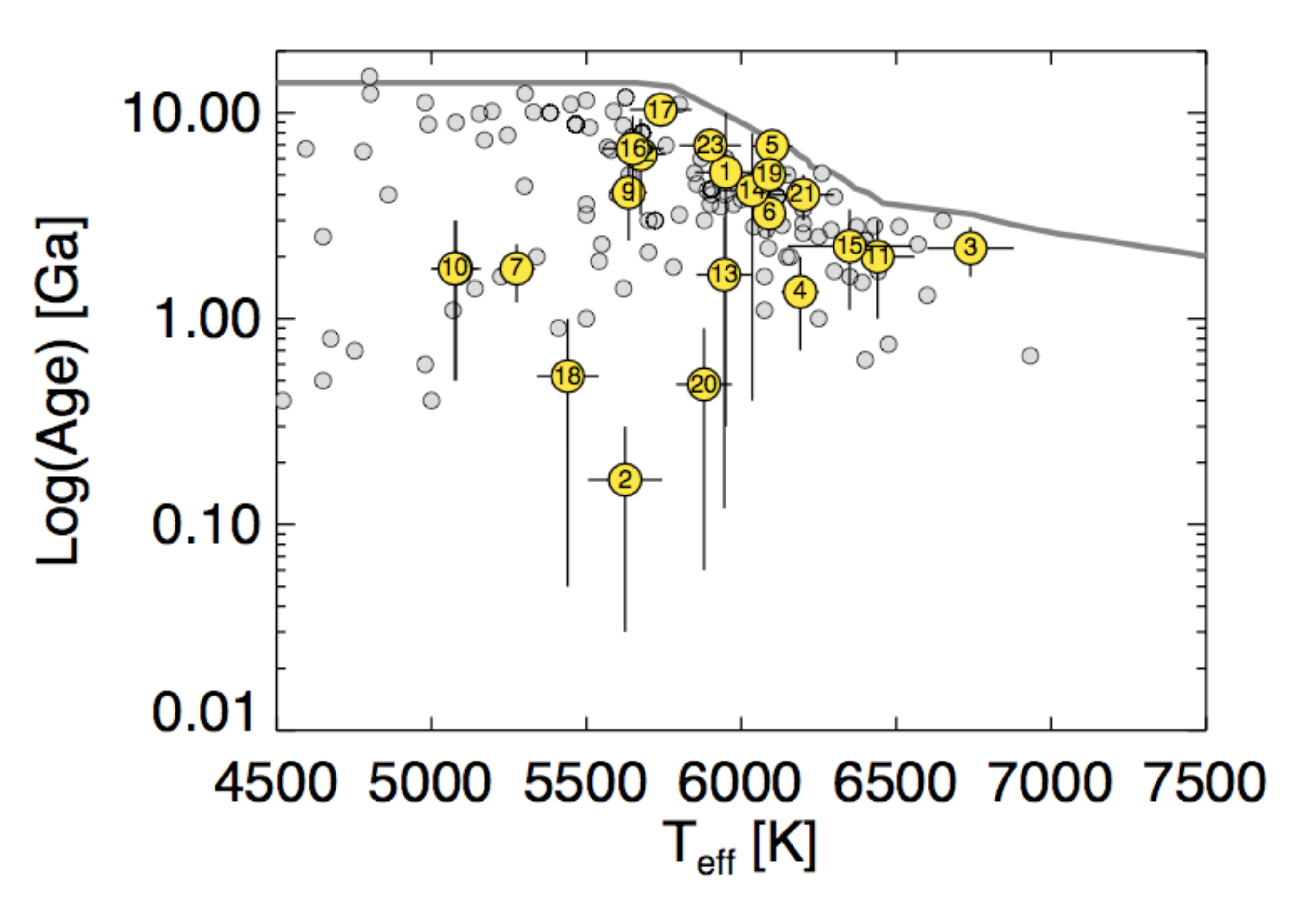}}
  \caption{Age of stars with transiting planets as a function of their effective temperature. The CoRoT planets are labelled 1 to 23. Ages of other transiting planets are obtained directly from the extrasolar planet encyclopedia (http://exoplanet.eu/). The grey line indicates the maximum main-sequence age of solar-composition stars as a function of their mid-life effective temperature \citep[from][]{morel}. Some of the CoRoT planets are very young and fill a void in that diagram.}
  \label{fig:justages}
\end{figure}

As shown in Figure~\ref{fig:justages}, three CoRoT stars with planets, CoRoT-2, 18 and 20 are estimated to be younger than 1\,Ga and probably only a few 100\,Ma. When accounting for the main-sequence lifetime of stars and the age of the Galaxy, this is consistent with a uniform age distribution. Over this effective temperature range, none of the other known transiting planets are that young. This can be explained by two factors: first most of the usual transit surveys are biased towards non-active stars both because of the photometric stability requirement for the transit detection and because of the easier radial-velocity follow-up. Second, the CoRoT systems have been studied more thoroughly than the average exoplanet, with in particular measurements of chromospheric activity, lithium abundance and stellar spin. These measurements are important to detect signs of youth which are generally not accessible from the study of the evolution tracks themselves.  

Two CoRoT systems are particularly interesting because of their complex lightcurves characteristic of spotted star, the fast stellar spin and presence of a close-in, relatively massive giant planet: CoRoT-2 and 18. Interestingly, these two systems are almost twins of each other. The two stars, CoRoT-2 \citep{alonso08} and CoRoT-18 \citep{hebrard11} have comparable effective temperatures (5450 vs. 5440 K), metallicities (0.0 vs. -0.1), spin periods (4.5 vs. 5.4 days) and $v\sin i$ (12 vs. 8.0$\rm\,km\,s^{-1}$), and they are both active, with peak to peak photometric variabilities of $\sim 4$\% and $\sim 2$\%, respectively. In addition CoRoT-2 and CoRoT-18 are the fastest-spinning stars with a known planet and an effective temperature below 6000\,K.

The two systems are indeed young: analysis of the X-ray activity of CoRoT-2 indicate an age of 100 to 300\,Myr \citep{schroter11}, whereas the spin of CoRoT-18 and its lithium abundance seem to indicate an age between 400 to 600\,Myr \citep{hebrard11}. The latter is problematic however, as the age obtained from evolution tracks would indicate an age in excess of 1\,Ga (or a star on the pre-main sequence, which is very unlikely). 

A third young system is CoRoT-20: the star's spin is not as important ($v\sin i=4.5\pm 1.0\,\rm km\,s^{-1}$), but it exhibits a clear lithium line, indicative of an age of $100^{+800}_{-40}$\,Myr \citep{deleuil20}. In this case, stellar evolution tracks only provide upper limits on the star's age (below 5\,Ga, within $1\sigma$). Without the lithium measurement, this star would probably have been attributed an older mean age compatible with a star on the main sequence. This examples illustrates how detailed follow-up studies are important in order to fully characterize these exoplanetary systems. Interestingly, this case also share similarities with the two others: the planet is massive ($4.24\pm 0.23\,\rm M_{\rm Jup}$), it has a larger orbital period ($P=9.24285$\,days) but a high eccentricity implies that it makes close approaches with the star and last but not least, its inferred size is also problematic. This time, it is too small, as shown by the very low radius anomaly and extremely high $M_Z$ value in table~\ref{tab:interior}. 

The fact that these three systems pose particular problems (the large size of CoRoT-2b, the inconsistent age determinations for CoRoT-18 and the small size of CoRoT-20\,b) may indicate that our knowledge of young, rapidly spinning, spotted stars is incomplete. The analysis of the CoRoT data combined with further follow-up observations should shed light on this important issue which bears directly on the problem of the formation of planetary systems. Some additional caution should be given to these analyses, since the spotted surface of the star has an expected impact on the true radius estimation \citep{czesla09,csizmadia13,torres}.
 
\subsection{The tidal interactions}

\citet{paetzold04} first showed that the combined measurement of stellar spin and orbital period of exoplanets could be used to constrain tidal dissipation inside stars and the fate of close-in planets. The detailed continuous lightcurves obtained by CoRoT are extremely useful to constrain precisely the stellar spin (when star spots are present) together with the characterization of the stellar and planetary parameters. 

CoRoT-3 and CoRoT-15 are two examples of systems in which a hot F star ($T_{\rm eff}=6740 \pm 140\,$K and $6350\pm 200\,$K) with massive companions ($21.8\pm 1.0$ and $63.3\pm 4.1\rm\,M_{Jup}$, respectively) are in or close to a so-called double-synchronization state: their orbital period ($4.2567994\pm 4.10^{-6}$ and $3.06036\pm 0.00003$ days) is in both case within the range of the measured spin periods. For CoRoT-3, the spin period was inferred from the vsini assuming $\sin i_*=1$ and give a range of 4 to 5 days. Interestingly \citet{triaud} find that the star and the orbit of the planet are misaligned by $\beta=37.6^{+10.0}_{-22.3}$ (with no constraint on $i_*$ however). This misalignment is tentatively explained by the elliptical instability estimated by \citet{cebronb}.
For CoRoT-15, the spin period was obtained directly from the lightcurve to be from 2.9 to 3.1 days. In both cases, this appears to suggest that the stars have been prevented to spin down or have been spun up by their massive companions. As discussed in \citet{bouchy,bouchy11b}, both CoRoT-3\,b and CoRoT-15\,b are archetypes of systems with massive, close-in companions which are found around F-type stars and not around G-type stars. The explanation probably lies in the fact that dissipation by internal gravity waves is not possible in these stars with a central convective zone \citep{barker10}, and that these stars also have a weaker magnetic braking, leading to a reduced loss of angular momentum. The combination of these two effects would lead to a prolonged life-time of the double-synchronisation state ($P_{\rm orb}=P_{\rm rot}$). 

One system appears to stand out in this respect: CoRoT-11 is yet another F star ($T_{\rm eff}=6440 \pm 140\,$K) with a very fast spin period $P_{\rm rot}=1.7\pm 0.2$\,days, orbited by a $2.33\,\rm M_{\rm Jup}$ planet with a period $P_{\rm orb}=2.99433$\,days. The star is spinning extremely rapidly, which is at odds with its relatively old inferred age (1 to 3\,Ga), and more importantly, the planet is spinning faster than its parent star, implying that the tidal torque pushes the planet outward instead of inward. \citet{lanza11c} propose that after the pre-main sequence phase, the star and planet may have been in near-synchronous rotation and that the star would have slowly pushed the planet outward since this time. In that case, the constraint on the stellar dissipation in the star would be $4\times 10^{6} < Q'_* < 2\times 10^7$, a relatively large value for F stars. Another possibility would be that the star-planet system have not evolved much tidally due to a much weaker dissipation, but this would imply that star must be much younger than inferred from its evolution tracks  \citep{gandolfi12}. Again, this is a problematic system involving a rapidly spinning star. 

Recently however, \citet{parviainen} found evidence for a secondary
eclipse of CoRoT-11 that implies a large eccentricity $e=0.35\pm 0.03$.  Although this should be checked by a combined fitting of the radial velocimetry and photometry data, this finding  changes the problem very significantly, both because of potential changes of the solution in terms of the planetary parameters, and because the system is now found to be in or close to pseudo-synchronization: The star has the same spin period as the planet's orbit {\it at periastron}. This pseudo-synchronisation period is \citep{hut81}:
\begin{equation}
P_{\rm pseudo}=P_{\rm orb}{\left(1+3e^2+{3\over 8} e^4\right)\left(1-e^2\right)^{3/2} \over 1+{15\over 2.d0} e^2+{45\over 8} e^4+{5\over 16}e^6}
\end{equation}
For CoRoT-11, we thus find $P_{\rm pseudo}=1.68\pm 0.09$\,days, perfectly compatible with the star's spin period. This system may thus be very similar in fact to CoRoT-3 and CoRoT-15.

Other CoRoT systems are interesting because of the combination of a short-period orbit and non-zero eccentricity, generally implying that the tidal dissipation in the planet may be constrained. The values are generally compatible with the global study of Hansen (2010) who found dissipation factors in exoplanets $10^7 < Q_{\rm p}' < 10^8$, or with values found for WASP-18 and WASP-19 \citep{brown11}.  For CoRoT-16b, $e=0.33\pm 0.10$ and $P_{\rm orb}=5.35227$\,days implies a relatively week dissipation $Q_{\rm p}' \sim 10^7$ to be compatible with the old stellar age \citep{ollivier}. For CoRoT-20b, the relatively large eccentricity appears to be damped only with a $7\,Ga$ timescale if $Q_{\rm p}' \sim 10^7$ and $Q'_*=10^7$ \citep{deleuil20}. 

Finally, CoRoT-21 is an interesting system in which the star is a sub-giant and the planet, with its close $2.72474\,$days orbit appears doomed to fall into the star in the next $800$\,Myr if the stellar dissipation is such that $Q'_* < 10^7$. 

As shown by these examples, the ensemble of CoRoT planets provides us with a rich variety of systems in which models of tidal interactions may be tested in great detail.

\subsection{Atmospheric properties of CoRoT planets}
Early discoveries of occultation signatures in CoRoT light curve have first been done by \citet{snellen,alonso09a,alonso09b} on CoRoT-1 b and CoRoT-2 b, both giant planets having a bright host star, a very short period, a large radius and a long sequence of data. The secondary eclipses have depths of about 160 and 60 ppm (resp. for CoRoT-1 b and 2 b) and emission temperatures of 2300 and 1900 K \citep{alonso09a,alonso09b}. In addition to the occultation, \citet{snellen} shows the phase curve of CoRoT-1 b as it orbits its star; this curve is characterised by a large temperature contrast between the dayside and the nightside. The redistribution of heat in the planet atmsophere is not efficient and  an upper value for the geometric albedo of 0.2 can be derived.

A recent study has applied an homogeneous processing to the CoRoT planet light curves in search for secondary transits \citep{parviainen}. The author announce three additional significant detections for CoRoT-6 b, CoRoT-11 b, and the brown dwarf CoRoT-15 b. Marginal detections are also reported for CoRoT-3 b, 13 b, 18 b, and 21 b. It allows refining eccentricity measurements and estimating brightness temperatures, which are systematically above the expected equilibrium temperatures. Additional heating mechanisms are required to explain the observations, and may be related to tidal effects (see previous section) or irradiation.
  
\subsection{Synergies between stellar and planetary studies}
The out-of-transit variations may give the possibility to derive the rotational properties of the host star: at minimum, its rotational period, and if the modulation is large and regular, the differential rotation and the location of active longitudes can be determined, using a starspot model (CoRoT-2, 4, 6, 7 by \citet{lanza09a,lanza09b,lanza10,lanza11a}). The shape of individual transits of CoRoT-2 b has also been used to estimate the properties of the spots while they are occulted by the planet (depth variations). As a consequence of the stellar activity, the radius of the planet is affected; for CoRoT-2 b, the planet radius was under-estimated by about 3\% \citep{silva}. In some cases, the spot evolution timescale may be related to the planetary motion, as if part of the activity was induced by star-planet interactions; it is observed when the planet orbit is faster, similar or slower than the stellar spin, respectively in the systems CoRoT-2, 4 and 6, and may be  explained by magnetic reconnections in the outer corona \citep{lanza11c}. 

The intrinsic stellar features sometimes interfere with the planetary signal that is searched for in the CoRoT light curves, especially for small-size planets. 
In the case of CoRoT-7, the precise modeling of the stellar superficial active regions is critical in order to correct for the radial-velocity jitter that superseeds the planet signal by a factor 5 at least \citep{lanza10,boisse,pont}. Depending on the modeling of the stellar modulation, the mass determination differs by a factor 2-3.

Finally, the transits of small planets can be used to measure the stellar limb darkening, which has an effect on the planetary size-determination, contributing to our knowledge about stellar atmospheres \citep{csizmadia13}.

\begin{landscape}
\begin{table*}
\caption{Physical and orbital characteristics of the extra-solar planets and brown dwarf discovered by CoRoT. The second line contains the errors.}
\label{planets}
\begin{center}
{\scriptsize
\begin{tabular}{lllllllllllllll}
\\
\hline
CoRoT- &	P     & Rp   & Mp  & e & a   & Ms    & Rs    & Teff & vsini& Fe/H & Prot  & Age & Ann. \\
Name    &days&Rjup&Mjup&    & AU&Msun&Rsun&K      &km/s &            &days&Gyr    &          \\
\hline
1b&1.5089557 & 1.49   & 1.03  &  0.006 &0.0254& 0.95 & 1.11  & 5950 & 5.2     &  -0.3   & 10.7 &  &\cite{barge}\\
     &0.00000064&0.08&0.12  &f0.012&0.0004& 0.15 &  0.05  & 150  &  1.0    &   0.25 &    2.2 &      & \\          
2b&1.7429964 & 1.47 & 3.31  & 0.036     &0.0281 & 0.97 & 0.90  & 5625 & 11.85 & -0.04 & 4.52 & 0.03-0.3 &\cite{alonso08}\\
     & 0.0000017 & 0.03& 0.16  &0.033&0.0009&0.06& 0.02 &  120  &  0.5 &   0.1 & 0.14   &      & \\          
3b&4.2567994     & 1.01    & 21.77 & 0.012 &0.057 & 1.37  & 1.56 & 6740 & 18     & -0.02  & 4.6  & 1.6-2.8 &\cite{deleuil3}\\
     & 0.000004& 0.07   &  1.0     &0.01&0.003  &  0.09 &  0.09&  140 &  3.0    &   0.06  & 0.4  &      & \\          
4b&9.20205     & 1.19    & 0.72 & 0.27  &0.090& 1.16 & 1.17 & 6190 & 6.4     & 0.05  & 8.9  & 0.7-2.0 &\cite{aigrain09}\\
     & 0.00037    & 0.06   &  0.08  & 0.15 &0.001&  0.03&  0.03 &  60    & 1.0     &  0.07 &  1.1 &      & \\          
5b&4.037896   & 1.33& 0.47  & 0.086 &0.0495& 1.00 & 1.19 & 6100 & 1.0  & -0.25 & 50 & 5.5-8.3 &\cite{rauer}\\
     & 0.000002 & 0.05 &  0.05 & 0.07 &0.0003&  0.02&  0.04 &65     &1.0   &   0.06 & 10  &      & \\          
6b&8.886593 & 1.17 & 2.96  & 0.18   &0.0855 &1.05 & 1.025& 6090 &  7.5  & -0.20 & 6.4 & 2.5-4.0 &\cite{fridlund}\\
     & 0.000004 & 0.04& 0.34  &  0.12 & 0.0015& 0.05 &  0.03&  50  &  1.0   &   0.1     & 0.5      & \\          
7b&0.853585   & 0.141 & 0.023& 0          &0.017& 0.91 & 0.82 & 5275 & 1.5   & 0.12  & 23.6& 1.2-2.3&\cite{leger}\\
     & 0.00005    & 0.009 &  0.0038& fixed  &0.001& 0.03 &  0.04& 60      & 1.0  & 0. 06 & 0.1&      & \\          
8b&6.21229   & 0.57 & 0.22   & 0   &0.063  & 0.88 & 0.77 & 5080 & 2.0     & 0.3    & 20 &0.5-3.0&\cite{borde}\\
     & 0.00003  & 0.02 &  0.03 &fixed&0.001&  0.04&  0.02 &  80    &  1.0   &   0.1    & 5   &      & \\          
9b&95.273804 & 0.94 & 0.84   & 0.11 &0.407& 0.99 & 0.94 & 5625 & 0.     & -0.01    & 14 & 0.5-8 &\cite{deeg09}\\
     & 0.0014    & 0.04   &  0.07   & 0.039 &0.005&0.04 & 0.04 &  80     &  1.0  &   0. 06   &   5 &      & \\          
10b&13.2406   & 0.97 & 2.75   & 0.53 &0.1055& 0.89 & 0.79 & 5075 &  2.    &  0.26    & 2.0 & 0.5-3.0&\cite{bonomo}\\
     & 0.0002    & 0.07   &  0.16  & 0.04 &0.0021&  0.05 &  0.05&  75    &  0.5  &   0.07  &  0.5  &      & \\          
11b&2.99433 & 1.43 & 2.33 &0.35       &0.0436&1.27 & 1.37 & 6440 &40.0 & -0.03  & 1.7 & 1-3 &\cite{gandolfi10}\\
     & 0.000011& 0.03 &0.34 &0.03  &0.005 & 0.05  & 0.03 &  120&  5.0&   0.08 &   0.2 &      & \\          
12b&2.828042& 1.44 & 0.92   & 0.07 &0.0402&1.08& 1.1& 5675 &1.0  & 0.16  & 68 & 3.2-9.4&\cite{gillon10b}\\
     & 0.000013 & 0.13 & 0.07   & 0.06 &0.0009&0.08&  0.1&  80    &  1.0 &   0.1          &   10 &      & \\          
13b&4.03519 & 0.89 & 1.31& 0.       &0.051&1.09   & 1.01  & 5945 &4    & 0.01   & 13 & 0.12-3.15&\cite{cabrera}\\
     & 0.00003  & 0.01 &0.07 & fixed  &0.0031&0.02&  0.03 &  90  &1.0   & 0.07   &   5 &      & \\          
14b&1.51214 & 1.09 & 7.6& 0.      &0.027&1.13 & 1.21  & 6035 &9.     & 0.05  & 5.7 & 0.4-8.0&\cite{tingley11a}\\
     & 0.00013  & 0.07 & 0.6 &fixed &0.002&0.09 &0.08   &  100  &0.5   & 0. 15 &   1 &      & \\          
16b&5.35227 & 1.17 & 0.54 & 0.33 &0.0618&1.1& 1.19 & 5650 &0.5   & 0.19   & 60 & 3.7-9.7&\cite{ollivier}\\
     & 0.0000    & 0.15 & 0.09 & 0.10 & 0.0015&0.08&0.14&  10  &  1.0  &   0.06 & 10   &      & \\          
17b&3.7681 & 1.02 & 2.43 &  0.     &0.0461&1.04 & 1.59& 5740 &4.5  & 0.00   & 20 & 9.7-11&\cite{csizmadia}\\
     & 0.0000    & 0.07& 0.30& fixed &0.0008& 0.1  &  0.07&  80    &0.5 &   0.1    &  5  &      & \\          
18b&1.900069 & 1.31 & 3.47  & 0.04  &0.0295&0.95 & 1.00 & 5440  & 8.0 & -0.10 & 5.4 & 0.05-1&\cite{hebrard11}\\
     & 0.0000    & 0.18   &  0.38&  0.04   &0.0016& 0.15& 0.13 &  100   & 1.0 &   0.1  & 0.4   &      & \\          
19b&3.89713 & 1.29 & 1.11   & 0.047&00518&1.21 & 1.65 & 6090 & 6.0     & -0.02   & 15&4.0-6.0&\cite{guenther}\\
     & 0.0000    & 0.03 &  0.06  & 0.045&0.0008&0.05 & 0.04&  70     &  1.0    &   0.1     &  5  &      & \\          
20b&9.24285 & 0.84 & 4.24   & 0.562 &0.0902&1.14 & 1.02  & 5880 &4.5   & 0.14    & 11.5 & 0.06-0.9&\cite{deleuil20}\\
     & 0.0000    & 0.04 &  0.23  & 0.013& 0.0021& 0.08 &  0.05 & 90     &1.0   & 0.12    &   3 &      & \\          
21b&2.72474 & 1.30 & 2.26   & 0  &0.0417  &1.29 & 1.95 & 6200&11.  & 0.00    & 10 & 3.0-5.0&\cite{paetzold12}\\
     & 0.0000    & 0.14 &  0.31 &fixed&0.0011&0.09 & 0.21&  100& 1.0   &   0.1     &   2 &      & \\          
23b&3.6313 & 1.05 & 2.8     & 0.16  &0.048&1.14 & 1.61 & 5900 &9.0    & 0.05     & 9.2 & 6.2-7.7&\cite{rouan}\\
     & 0.0001  & 0.13 & 0.3     &  0.02 &0.004& 0.08& 0.18 &  100  &1.0    &   0.1      &  1.5  &      & \\          
\hline
15b&3.06036 &1.12   & 63.3    & 0       &0.045&1.32& 1.46 & 6350 & 19 & 0.1    &3.0  &1.1-3.4 &\cite{bouchy}\\
     & 0.00003  & -0.15,0.3&4.1&fixed &0.014&0.12&-0.14,0.31&200 &1.0 & 0.2  &  0.1  &      & \\          
\hline
\\
\end{tabular}  }
\end{center}
\end{table*}
\end{landscape}

\section{Conclusion and future}

In addition to planet detection and characterization, the scientific results achieved in the exoplanet fields observed by CoRoT are enriched by the variety of available targets, and by a series of complementary observations from ground-based observatories. The highlights of this additional science are: i) the first mass measurement of a brown dwarf using Doppler boosting in the CoRoT-3 system \citep{mazehc3}, ii) the first asteroseismic analysis of red giants \citep{kallinger,hekker,mosser}, iii) the characterization of the stellar population \citep{gazzano10,sebastian,guenther}, iv) the classification of stellar variability \citep{debosscher}, v) the search for transit timing variations in the light curves of CoRoT-1\,b and CoRoT-9\,b \citep{csizmadia10,borkovits}, vi) the measurement of spectroscopic transits for estimating the orbital alignment of  the planet with respect to the stellar spin axis in the systems CoRoT-1, 2, 3 and 11 \citep{pont10,bouchy08,czesla12,triaud,gandolfi12}...

The content of the CoRoT light curve archive will remain a long-term source of discoveries, both in stellar physics and in the field of exoplanet science. With a largest community of users searching the data, and with the  evolution of analysis tools, one may expect the detection of new transiting events, occultations of planets or thermal/reflected light. The CoRoT data pipeline is constantly improved (background correction, hot pixel identification and removal, contamination and jitter calibration) and future data products will eventually appear in the archive. The vast community of transit hunters is welcome to use all public data, released in http://idoc-corotn2-public.ias.u-psud.fr. In addition to transit detection, complementary observations are also encouraged on CoRoT candidates, in particular to: i) provide high angular-resolution images of candidates, and high-contrast detection limits in their close vicinity (as in \citet{guenther12}), ii) ground-based transit observations using 1m-class telescopes, to maintain the ephemeris accuracy over long periods of time, iii) spectroscopy of host planets for a more accurate determination of their fundamental parameters (metallicity, age, radius), iv) high-precision radial-velocity measurements when necessary, to better constrain the mass and eccentricity of the planets and search for additional planetary companions in the system. The prospect to solve the remaining promising CoRoT planet candidates thus depends on the use of the legacy archive. All available existing complementary data on candidates that have been followed up in the framework of the CoRoT project are commented in \citet{mega}.\\

At the time of writing, the CoRoT satellite encounters severe electrical problems and the observations have been discontinued on November 2nd, 2012 \footnote{see Nature doi:10.1038/nature.2012.11845}. If the failure can be repaired, the scientific program will be carried on during a second period of extension that could last up to March 2016. In particular, the ability of CoRoT to point any direction within its two "eyes" will be used to further explore the bright stellar content. In the exoplanet channel, the focus will be put on stars already known to host extrasolar systems, in the objective of searching for their transit, for the transit of inner planets in the system, or for the signature of the reflective phase curve if the system is not  transiting and period is short. A marginal detection of this reflection curve has been done on HD~46375 \citep{gaulme}, and this star is observed again during the current season (October 2012 to March 2013), in particular to confirm this previous marginal detection. Other planetary systems detected by their radial velocity signature, like the super-earth HD~179079\,b, or the system HD~52265\,b have been observed by CoRoT \citep{escobar}. Note that the pointing limits of CoRoT from the center of its eyes have been recently relaxed to a circle of 14 degrees radius, due to the very efficient baffling performance.

The future of CoRoT is uncertain now, as presently the scientific data cannot be collected due to a failure of the electronics of the instrument. The satellite is working well and the instrument is in a safe mode. At the end of the exploitation period, the satellite would be slowed down to decreasing orbital distances until its long-term return to Earth. Kepler presently continues observing its unique stellar field in Cygnus up to 2016. The future of space missions towards transiting exoplanets is rapidly evolving since 2011, and unclear at present. Several projects are under study at ESA and NASA, aiming at either discovering transits of Earth-like planets around the brightest stars (PLATO, TESS, CHEOPS), or at characterizing the atmosphere of a subsample of transiting planets (ECHO, FINESSE). Meanwhile, ground-based efforts are carried on to improve the precision of small-telescopes networks with HAT-South and the Next Generation Transit Survey, in the intermediate range of stellar magnitudes and more especially focussed on low-mass stars. In combination with the ongoing and uprising radial-velocity surveys optimised for low-mass planets, this is a very promising future for comparative exoplanetology that arises.

\section*{Acknowledgements}
The CoRoT Exoplanet Science Team is very grateful to the CNES for operating CoRoT for six years and supporting science activities on CoRoT data.
Besides of the authors of this paper, the CoRoT Exoplanet Team is composed of: J.M. Almenara, M. Auvergne, S. Barros, A. Bonomo, C. Damiani, R. D\'{i}az, G. H\'ebrard, A. L\'eger, G. Montagnier, M. Ollivier, D. Rouan, A. Santerne, J. Schneider (F) A. Erikson, E. Guenther, A. Hatzes, M. P\"atzold, A. Ofir, H. Rauer, G. W\"uchterl (G), R. Alonso, H. Parviainen, B. Tingley (S), S. Carpano, D. Gandolfi, M. Fridlund (ESA), R. Dvorak, H. Lammer, J. Weingrill (A), T. Mazeh, L. Tal-Or (Is), S. Ferraz-Mello (Br), S. Aigrain (UK), M. Gillon (B), M. Endl (US).
HD acknowledges support by
grant AYA2010-20982-C02-02 and AYA2012-39346-C02-02 of the Spanish
Ministerio de Economi\'a y Competividad.

\bibliographystyle{elsarticle-harv}

\end{document}